\documentclass[journal]{IEEEtran}
\IEEEoverridecommandlockouts
\usepackage{amsmath,amssymb,amsfonts}
\usepackage{algorithmic}
\usepackage{graphicx}

\usepackage{accents}
\usepackage{textcomp}
\def\BibTeX{{\rm B\kern-.05em{\sc i\kern-.025em b}\kern-.08em
    T\kern-.1667em\lower.7ex\hbox{E}\kern-.125emX}}

\usepackage{lipsum} % print dummy text
 \usepackage{balance}
\usepackage{booktabs} % required for showing good \hline in matrix form --> \midrule
\usepackage{enumitem}
\usepackage{mathtools}
\usepackage[lined,boxed,commentsnumbered,linesnumbered,ruled]{algorithm2e}
\usepackage{tikz} % common graphical tools for pictures
\usetikzlibrary{shapes, patterns,decorations.text, decorations.pathreplacing}
\usetikzlibrary{external}

\usepackage{soul}

\usepackage{array,verbatim}
\usepackage{makecell}
\usepackage{ctable}
\usepackage{subcaption}
\usepackage[noadjust]{cite} % Improves presentation of citations, like [3--7].
\usepackage{pgfplots}
\pgfplotsset{filter discard warning=false}
\pgfplotsset{compat=1.14}
\usepgfplotslibrary{colorbrewer}
\usepgfplotslibrary{groupplots}
\usetikzlibrary{intersections, pgfplots.fillbetween}
\usetikzlibrary{arrows}

\usepackage[capitalize]{cleveref}
\crefname{equation}{\unskip}{\unskip}
\crefname{claim}{Claim}{Claims} %\crefname{type}{singular}{plural}
\usepackage{array}
\usepackage{multirow}
\newcolumntype{C}[1]{>{\centering\arraybackslash}p{#1}}
\allowdisplaybreaks

\newcommand{\Li}{\mathcal{L}}
\newcommand{\Di}{\mathcal{D}}

\newcommand{\Pt}{\mathcal{P}}
\newcommand{\Ls}{\widetilde{L}}

\newcommand{\estar}{\epsilon^*}

\newcommand{\e}{\epsilon}

\newcommand{\defn}{\triangleq}

\newcommand{\BERtw}{P_\mathsf{b,t,sw}}

\newcommand{\BERs}{P_\mathsf{b,str}}

\newcommand{\BLERtw}{P_\mathsf{bl,t,sw}}

\newcommand{\BLERs}{P_\mathsf{bl,str}}
\newcommand{\nusingle}{\breve{\nu}}

\newcommand{\dv}{d_\mathsf{v}}
\newcommand{\dc}{d_\mathsf{c}}

\newcommand{\dip}{\delta}
\newcommand{\dipc}{\kappa}
\newcommand{\pd}{\psi_{\mathsf{d}}}
\newcommand{\estard}{\estar_{\mathsf{d}}}
\newcommand{\estarsc}{\estar_{\mathsf{SC}}}
\newcommand{\estarunc}{\estar_{\mathsf{unc}}}

\providecommand{\keywords}[1]
{
    \vspace{-5pt}
    {\small	\textbf{\textit{Index Terms---}#1.}}
}

\begin{document}

\title{On Doped SC-LDPC Codes for Streaming}
\author{
 Roman Sokolovskii, \IEEEmembership{Graduate Student Member, IEEE}, Alexandre Graell i Amat, \IEEEmembership{Senior Member, IEEE}, \\and Fredrik Br\"annstr\"om, \IEEEmembership{Member, IEEE}
 \vspace{-9pt}
     \thanks{This work was funded by the Swedish Research Council (grant 2016-4026).}
     \thanks{R. Sokolovskii, A. Graell i Amat, and F. Br\"annstr\"om are with the Department of
     Electrical Engineering, Chalmers University of Technology, SE--41296 Gothenburg, Sweden (email:
     \{roman.sokolovskii,alexandre.graell,fredrik.brannstrom\}@chalmers.se).}}

\maketitle

\begin{abstract}
    In streaming applications, doping improves the performance of spatially-coupled low-density parity-check (SC-LDPC) codes by creating reduced-degree check nodes in the coupled chain.
    We formulate a scaling law to predict the bit and block error rate of periodically-doped semi-infinite SC-LDPC code ensembles streamed over the binary erasure channel under sliding window decoding for a given finite component block length.
    The scaling law assumes that with some probability doping is equivalent to full termination and triggers two decoding waves; otherwise, decoding performs as if the coupled chain had not been doped at all.
    We approximate that probability and use the derived scaling laws to predict the error rates of SC-LDPC code ensembles in the presence of doping.
    The proposed scaling law provides accurate error rate predictions.
    We further use it to show that in streaming applications periodic doping can yield higher rates than periodic full termination for the same error-correcting performance.
\end{abstract}
\keywords{Finite-length code performance, spatially-coupled LDPC codes, window decoding, streaming}

\section{Introduction}
\label{sec:intro}

Spatially-coupled low-density parity-check (SC-LDPC) \mbox{codes~\cite{ref:Jime99,ref:Lent10}} exhibit \textit{threshold saturation}---in the limit of large block length, the error-correcting performance of suboptimal belief propagation (BP) decoding approaches that of optimal maximum a posteriori decoding of the underlying uncoupled ensemble.
First observed numerically~\cite{ref:Lent10}, threshold saturation was proven for the binary erasure channel (BEC)~\cite{ref:Kude11} and later for any binary-input memoryless symmetric channel~\cite{ref:Kude13}.
Moreover, the minimum distance of regular SC-LDPC code ensembles grows linearly with the block length.

A spatially-coupled code is constructed from a sequence of Tanner graphs by interconnecting them according to a predefined pattern.
To yield threshold saturation, the resulting coupled chain must contain spatial positions where bits are better protected than in the original (uncoupled) code.
This may be achieved by terminating the chain, which guarantees lower average degrees of the check nodes (CNs) at the termination boundaries.
During BP decoding, more reliable information propagates from the termination boundaries inward in a wave-like fashion.
These decoding waves are not guaranteed to propagate through the whole chain---the longer the chain, the more likely it is that they get stuck and a decoding error occurs.
This phenomenon is often referred to as \textit{decoding error propagation,} especially in the context of sliding window (SW) decoding and streaming applications.
Thus, to limit decoding error propagation, one would wish to keep the chain length low.
Short chains, however, entail significant rate loss that hinders efficient data transmission.

To limit the rate loss and at the same time alleviate performance degradation associated with long coupled chains, Zhu \textit{et al.}~\cite{ref:Zhu20} proposed to occasionally insert spatial positions that contain reduced-degree CNs---a technique called \textit{CN doping}.
Alternatively, \textit{variable node (VN) doping}~\cite{ref:Zhu20_vn}, where all VNs at some spatial positions are fixed, may be considered.
A more general technique was introduced in~\cite{ref:Camm16}, where a fraction (i.e., not necessarily all) of VNs at a doped position are fixed.

Doping is especially potent in streaming applications, where the risk of decoding error propagation is most substantial~\cite{ref:Zhu20_vn}.
It allows to flexibly trade off code rate for improved finite-length performance.
(We use the term \textit{finite-length} to refer to finite component block length---the length of the coupled chain may still be infinite.)
To efficiently navigate the space of possible trade-offs, it is of significant practical importance to develop a model that accurately predicts the finite-length performance of doped codes.
In this paper, we derive a finite-length scaling law for the bit error rate (BER) and block error rate (BLER) of VN-doped SC-LDPC code ensembles transmitted in a streaming fashion over the BEC under SW decoding~\cite{ref:Iyen12}.
In particular, we extend the scaling law for terminated SC-LDPC code ensembles~\cite{ref:Soko19,ref:Soko20} to semi-infinite chains with VN doping at regular intervals.
To that end, we first consider terminated ensembles with a single doping point and provide a finite-length scaling law that characterizes the probability that the doping point triggers the decoding waves.
We also apply this model to predict the finite-length performance of the more general doping in~\cite{ref:Camm16}. 
We use the derived finite-length scaling law to find the longest doping interval that guarantees a given target BER and demonstrate that in streaming applications doping can yield higher design rates than those achievable with periodic full termination.

\section{Preliminaries}
\label{sec:preliminaries}

We consider VN doping of semi-structured $(\dv,\dc,L,N)$ SC-LDPC code ensembles.
For the purposes of the analysis, we first consider tail-biting and terminated SC-LDPC codes and then apply gained insights to the streaming scenario and semi-infinite chains.
A semi-structured SC-LDPC code~\cite{ref:Olmo15} is composed of $L$ $(\dv,\dc)$-regular LDPC codes, each of length $N$, arranged in a sequence of spatial positions indexed by $i \in \Li = \{0,\ldots,L-1\}$.
The edges of the corresponding Tanner graph are then permuted to guarantee the following connectivity: every VN at position $i \in \Li$ is connected to a CN (chosen uniformly at random) at $\dv$ consecutive positions in the range $[i,\ldots,i+\dv-1]$.
If the chain is terminated, $\dv-1$ positions containing VNs only are appended at the end of the chain to connect the overhanging edges.
In contrast to VNs, no particular spatial connectivity is enforced upon CNs in the resulting Tanner graph, i.e., a CN at position $i \in \{ \dv - 1, \ldots, L - 1 \}$ is connected to $\dc$ VNs from any non-empty subset of positions in the range $[i-\dv+1,\ldots,i-1,i]$.
A detailed procedure to generate elements from the terminated semi-structured $(\dv,\dc,L,N)$ SC-LDPC code ensemble is described in~\cite{ref:Olmo15}.
This ensemble is akin to protograph-based  ensembles from the VN perspective; from the CN perspective, however, it resembles an ensemble with \textit{smoothing}~\cite{ref:Kude11}.
Although of little practical importance, this ``semi-structured'' construction was proposed in~\cite{ref:Olmo15} to simplify the finite-length analysis.
The analytical tools originally developed in~\cite{ref:Olmo15} for semi-structured ensembles were later successfully adapted to ensembles based on protographs~\cite{ref:Stin16}. The analysis in this paper can also be extended to protograph-based constructions.

\begin{figure}[!t]
\vspace{-1ex}
    \centering
    \includegraphics[width=\columnwidth]{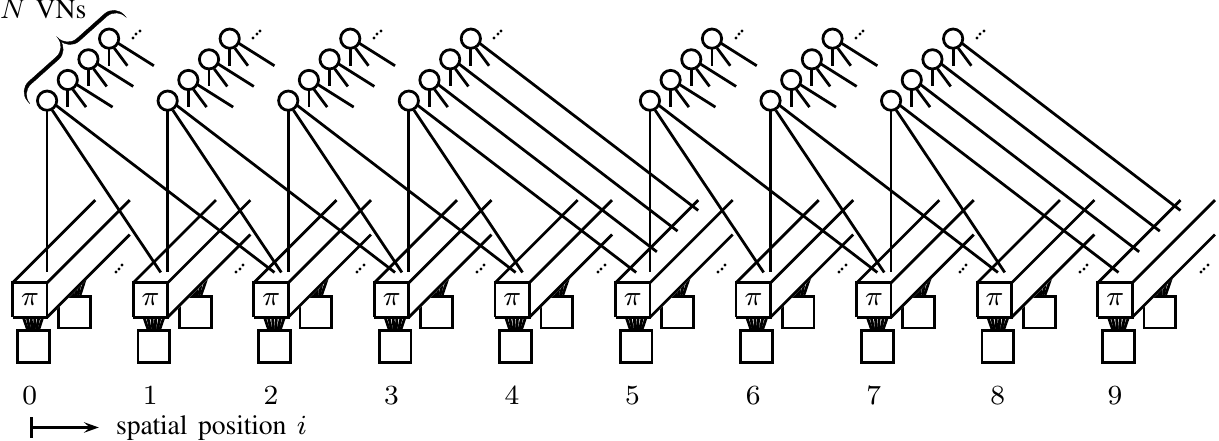}
    \caption{Tanner graph of the doped terminated $(\dv,\dc,L,\Pt,N)$ SC-LDPC code ensemble with $\dv=3,\dc=6,$ $L=8,$ and $\Di = \{ 4 \}$.}
\label{fig:doped_ensemble}
\vspace{-3ex}
\end{figure}

We analyze two schemes that allow the code designer to flexibly trade off code rate for improved finite-length performance: VN doping~\cite{ref:Zhu20_vn} and the more general doping in~\cite{ref:Camm16}.
VN doping consists of fixing all VNs in a subset of positions $\Di \subseteq \Li$ to zero.
Note that fixing $\dv-1$ consecutive positions is equivalent to full termination.
Following~\cite{ref:Camm16}, we also study a more general version of VN doping where a fraction $0 < \alpha_i \leq 1$ of VNs at a doped position $i \in \Di$ are fixed.
(The case $\alpha_i = 1~\forall~i \in \Di$ corresponds to the VN doping in~\cite{ref:Zhu20_vn}.)
We refer to this more general setup as \textit{soft} VN doping, as opposed to \textit{hard} VN doping~\cite{ref:Zhu20_vn}.
We use the term \textit{doping point} to refer to the location in the chain where doping is applied.
In contrast, we refer to a specific way to apply doping as a \textit{doping pattern} $ \Pt \defn (\Di,\boldsymbol{\alpha})$, where $\boldsymbol{\alpha}$ is a vector indexed by elements from $\Di$.
Throughout the paper, we assume hard doping unless $\boldsymbol{\alpha}$ is explicitly specified.
We note that in the context of terminated or tail-biting $(\dv,\dc,L,\Pt,N)$ SC-LDPC codes with doping, $L$ includes $|\Di|$ doped positions.
Fig.~\ref{fig:doped_ensemble} shows the Tanner graph of a terminated semi-structured ensemble with doping at position 4 for illustrative purposes.

For streaming-oriented applications, we consider the transmission of semi-infinite $(\dv,\dc,\Ls,\Pt,N)$ SC-LDPC code chains with doping at regular intervals.
In this context, $\Ls$ refers to the number of spatial positions between consecutive doping points, and $\Di$ specifies doping with reference to each doping point.
For example, $\Ls=50$ and $\Di= \{ 0, 1, 2 \}$ specify a semi-infinite chain where the sequence of doped positions is $\{ 50, 51, 52, 103, 104, 105, \ldots \}$.

We consider transmission over the BEC with erasure probability $\e$ and decoding via peeling decoding.
During the initialization of peeling decoding, all non-erased VNs are deleted from the Tanner graph of the code along with adjacent edges.
Subsequently, at each iteration, a degree-one CN is randomly chosen, the bit associated with the neighbor VN is recovered, and the VN is removed from the graph along with the $\dv$ edges connected to it.
Decoding fails if the decoder runs out of degree-one CNs before recovering all erased VNs.
If the number of iterations is not limited, the performance of peeling decoding is equivalent to that of BP decoding for the BEC.
Using statistical properties of the stochastic process associated with the number of degree-one CNs over the iterations of peeling decoding, the scaling laws in~\cite{ref:Amra09,ref:Olmo15,ref:Soko20} estimate the probability that the number of unrecovered VNs at the end of decoding is linearly sized with respect to the component code length $N$, thereby predicting the frame error rate (FER) of the uncoupled~\cite{ref:Amra09} and spatially-coupled~\cite{ref:Olmo15,ref:Soko20} ensembles in the waterfall region.

As in~\cite{ref:Soko20}, here we study the performance of the practical SW decoding~\cite{ref:Iyen12}, which restricts BP message passing to CNs in a window of $W$ spatial positions.
After a specified number of BP iterations, the decoder makes a decision on the $N$ bits in the leftmost position within the window, and the window slides by one position to the right.
SW decoding yields decoding latency of $N(W+\dv-1)$ bits, as opposed to $NL$ bits for full BP decoding.
However, the SW restricts one of the two decoding waves in the terminated chain to the last $W$ positions~\cite{ref:Soko20}, yielding higher error rates.

\section{Density and Mean Evolution}
\label{sec:fl_scaling}

To estimate the finite-length performance of doped ensembles, we first need to understand their asymptotic behavior.
To that end, we employ density evolution to compute the iterative decoding threshold $\estard$ for a given doping pattern~$(\Di,\boldsymbol{\alpha})$.
\vspace{-5pt}

\subsection{Decoding Thresholds via Density Evolution}
\label{sse:thresholds}

Some care should be exercised when choosing the initial conditions of density evolution for SC-LDPC code ensembles with doping: we need to isolate the effect of doping on the iterative threshold from that of the termination boundaries.
It is customary to achieve this by means of tail-biting---tail-biting SC-LDPC code ensembles do not have any termination boundaries, so the only source of structured irregularity in the spatially-coupled chain is the doped positions.

Let $q_{i+j,i}^{(\ell)}$ denote the probability that a CN at position $i+j$ will send an erasure message to a VN at position $i$ at BP iteration $\ell$.
Likewise, let $p_{i,i+j}^{(\ell)}$ be the probability that a VN at position $i$ will send an erasure message to a CN at position $i+j$ at BP iteration $\ell$.
Henceforth, due to tail-biting, index arithmetic must be performed modulo $L$.
The CN update for the semi-structured ensemble uses the property that a VN connected to a CN at position $i\in\Li$ is located at a position uniformly distributed in the range $[i-\dv+1,\ldots,i-1,i]$.
Hence, to get the outgoing CN error probability, we average the incoming error probabilities and obtain
\begin{equation}
    q_{i+j, i}^{(\ell)} = 1 - \left( 1 - \frac{1}{\dv}\sum_{j'=0}^{\dv-1}  p_{i+j-j',i+j}^{(\ell-1)} \right)^{\dc - 1}.
\label{eq:cnupd_semi}
\end{equation}
Since a VN at position $i\in\Li$ is connected to $\dv$ consecutive positions $\{i, i+1, i+\dv-1\}$, as we discussed in Section~\ref{sec:preliminaries}, the VN update for the semi-structured ensemble is
\begin{equation}
p_{i, i+j}^{(\ell)}=\e \left( 1 - \alpha_i \right) \prod_{j^{\prime} \neq j} q_{i+j^{\prime}, i}^{(\ell)}.
\label{eq:vnupd}
\end{equation}
To numerically estimate the decoding threshold $\estard$ associated with a certain doping pattern, we initialize $p_{i,i+j}^{(0)} = 1$ for all $i \in \Li, j \in \{0,\ldots,\dv-1\}$ and iterate equations~\eqref{eq:cnupd_semi}--\eqref{eq:vnupd} until the average VN erasure probability converges either to zero (for $\e \le \estard$) or to a fixed point (for $\e > \estard$).

\newcommand{\tablehighlight}[1]{\textbf{#1}}
\begin{table}[t]
    \caption{BP decoding thresholds for doped tail-biting semi-structured $(5,10,L,\Pt,N)$ SC-LDPC code ensembles. Hard doping is assumed unless $\boldsymbol{\alpha}$ is specified. The parameter $\dipc$ is introduced in Section~\ref{sec:scaling}. }
	\centering
	\renewcommand{\arraystretch}{1.2}
	\begin{tabular}{cccc}
		\toprule
        \tablehighlight{Alias} & \tablehighlight{Doping Pattern} & \tablehighlight{BP threshold $\estard$} & \tablehighlight{$\dipc$} \\
		\midrule
        ---                     & $\emptyset$        & 0.3415 & ---     \\
        ---                     & $\{ 0 \}$          & 0.3743 & 7.5444  \\
        ---                     & $\{ 0, 1 \}$       & 0.4244 & 3.4410  \\
        $\Pt_3$                 & $\{ 0, 1, 2 \}$    & 0.4783 & 2.5044  \\
        ---                     & $\{ 0, 1, 2, 3 \}$ & 0.4994 & ---     \\
        $\Pt_{\mathsf{spaced}}$ & $\{ 0, 2, 4 \}$    & 0.4979 & 2.1843  \\
        \midrule
        $\Pt_{\mathsf{soft}}$ &
        \begin{math}
            \begin{aligned}
                \Di &= \{ 0, 1, 2, 3, 4 \} \\
                \boldsymbol{\alpha} &= \left(0.75,0.2,0.75,0.2,0.75\right)
            \end{aligned}
        \end{math}
                & 0.4688  & 2.5067 \\
		\bottomrule
	\end{tabular}
	\label{tab:thresholds}
\vspace{-3ex}
\end{table}

Table~\ref{tab:thresholds} shows the BP thresholds for several doped tail-biting semi-structured $(5,10,L,\Pt,N)$ SC-LDPC code ensembles with different doping patterns.
If no positions are doped, the decoding threshold coincides with the BP threshold of the underlying uncoupled $(5,10)$-regular LDPC code ensemble, $\estarunc \approx 0.3415$.
If four consecutive positions are doped, which corresponds to full termination, the decoding threshold is equal to the threshold of the terminated ensemble, $\estarsc \approx 0.4994$.
The thresholds for one to three consecutive doped positions lie between these two extremes.
Interestingly, \textit{spacing} the three doped positions greatly improves the BP threshold.
Indeed, the threshold for three spaced doped positions (the penultimate row of Table~\ref{tab:thresholds}), $\estard\approx 0.4979$, is very close to that for full termination, $\estarsc\approx 0.4994$.

We also observe that soft doping enables us to flexibly trade off asymptotic performance for smaller rate loss---with $\sum_i \alpha_i / L = 2.65 / L$ of the VNs fixed, the soft doping pattern shown in the last row of Table~\ref{tab:thresholds} achieves a threshold that lies between those for two and three consecutive doped positions, where $2/L$ and $3/L$ of the VNs are fixed, respectively.
\vspace{-5pt}

\subsection{Mean Evolution and Decoding Trajectories}
\label{sse:mevol}

\begin{figure}[!t]
\vspace{-1ex}
    \centering
    \includegraphics[width=\columnwidth]{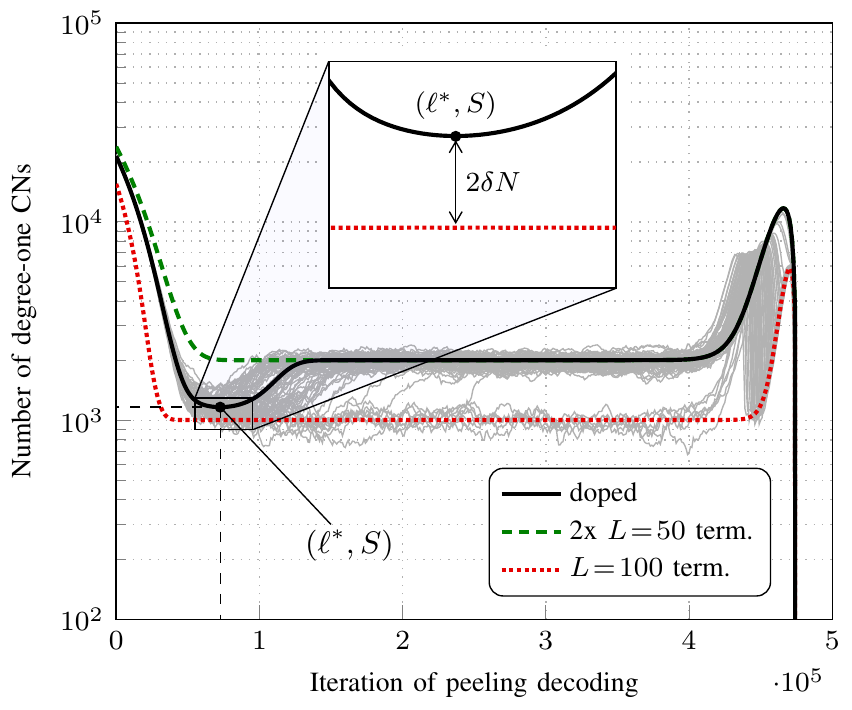}
    \caption{Mean evolution for the doped terminated $(5,10,L\!=\!103,\Pt,N\!=\!10^4)$ SC-LDPC code ensemble with $\Di = \{ 50, 51, 52 \}$ (black solid) at $\e = 0.475$.}
\label{fig:mevol_0475}
\vspace{-3ex}
\end{figure}

Mean evolution tracks the average number of degree-one CNs available to the peeling decoder through iterations via a system of coupled differential equations.
We employ the same mean evolution equations as in~\cite{ref:Olmo15}.
To take doping into account, we must merely update the initial conditions.
The details are trivial and are omitted.
Fig.\,\ref{fig:mevol_0475} shows the evolution of the average number of degree-one CNs during the iterations of the peeling decoding for the terminated $(5,10,L\!=\!103,\Pt,N\!=\!10^4)$ SC-LDPC code ensemble with three doped positions in the middle of the chain, $\Di = \{ 50, 51, 52 \},$ for $\e=0.475$ (black solid).
The erasure probability $\e=0.475$ is below the decoding threshold for this doping pattern, $\estard\approx0.4783$.
The steady state phase of the mean evolution below the threshold follows the mean evolution for two fully terminated chains of length $L\!=\!50$ (green dashed) after a transition period.
This corresponds to the presence of four decoding waves: two waves emanating from the termination boundaries at the edges of the chain and two more from the doping point in the middle.
The two waves from the doping point do not form steadily, as do those from full termination.
Instead, these waves subside to a local minimum before regrowing and starting to propagate along the coupled chain, a behavior central to our analysis.

Moreover, doping does not always result in two additional decoding waves in the mean evolution.
If we set $\e$ above the threshold $\estard$, the black solid curve will follow the mean evolution for a single fully terminated chain of length $L\!=\!100$ (red dotted), indicating that no decoding waves are triggered by doping in the mean evolution for $\e > \estard$.

Fig.~\ref{fig:mevol_0475} also shows several simulated realizations of the number of degree-one CNs available for peeling decoding as a function of the decoding iteration (thin gray solid curves), where for each realization an element of the ensemble is drawn at random.
We observe that for $\e=0.475$, an erasure probability that is below the threshold $\estard\approx0.4783$, most of the trajectories follow the black solid mean evolution curve and are characterized by four decoding waves.
Some trajectories, however, fail to transition to the four-wave regime and follow the red dotted curve of the two-wave process instead.
This suggests that doping is not guaranteed to work even if the system operates below the threshold $\estard$.
Moreover, note that the trajectories bifurcate to either four- or two-wave steady-state level.
This means doping is an all-or-nothing phenomenon: it either produces two decoding waves from the doping points or none at all.

\section{Finite-Length Scaling Law}
\label{sec:scaling}

The observations in the previous section motivate us to propose a finite-length scaling model wherein a doping point is equivalent to a probabilistic ``switch'' between full termination and an unobstructed chain.
We will first carry on with our running example of terminated ensembles with a single doping point and then consider semi-infinite chains with regular doping.
As we have seen in Section~\ref{sse:mevol}, mean evolution of terminated ensembles with doping has a local minimum before the steady state phase.
Let $S$ denote the number of degree-one CNs at that minimum and $\ell^*$ the iteration of peeling decoding when this minimum is achieved (marked on the black solid curve in Fig.~\ref{fig:mevol_0475}).
At this point, the two waves that emerge from the termination ends have already formed.
The number of degree-one CNs in these two waves is approximately $\gamma \left( \estarsc - \e \right)N$, where $\gamma$ is an ensemble-dependent constant~\cite{ref:Olmo15}.
In Fig.~\ref{fig:mevol_0475}, this number corresponds to the steady-state value of the mean evolution for the terminated ensemble with $L=100$ (red dotted).
The remaining degree-one CNs are located in the neighborhood of the doping point; for symmetric doping patterns, to which we limit the scope of this paper, they are equally distributed on each side of the doping point.
Let $\dip$ denote the number of degree-one CNs at one side of the doping point at iteration $\ell^*$, normalized by~$N$,
\begin{equation}
    \dip \defn S / (2N) - \gamma \left( \estarsc - \e \right) / 2.
    \label{eq:dip}
\end{equation}
In other words, up to normalization by $N$, $\dip$ denotes half the distance between the red dotted and the black solid curve in Fig.~\ref{fig:mevol_0475} at the local minimum~(shown in the zoomed-in section).

We empirically observe that $\dip(\e)$ is closely approximated by a line that hits zero at the threshold $\estard$,
\begin{equation}
    \dip(\e) \approx \dipc \left( \estard - \e \right).
    \label{eq:dipline}
\end{equation}
We treat $\dipc$ as a scaling parameter that depends on the underlying LDPC code ensemble and the doping pattern, and estimate it from numerical values of $\dip(\e)$ for several $\e$.
For our running example of three consecutive doped positions in the $(5,10,L,\Pt,N)$ SC-LDPC code ensemble, $\dipc \approx 2.5044$.
The corresponding values for some other doping patterns are reported in the last column of Table~\ref{tab:thresholds}.

We assume that a doping point produces two decoding waves with probability $\pd$.
Since, according to~\cite{ref:Soko20,ref:Amra09,ref:Olmo15}, the decoding process associated with the number of degree-one CNs present in each of the decoding waves over iterations converges to a Gaussian process with variance $\nusingle / N$---where $\nusingle$ is an ensemble-dependent constant that we estimate as in~\cite{ref:Soko20} from the truncated ensemble---we approximate $\pd$ with the probability that a Gaussian random variable with mean $\dip$ and variance $\nusingle / N$ is positive,
\begin{equation}
    \pd \approx 1 - \mathsf{Q}\left( \frac{\dipc \left( \estard - \e \right)}{\sqrt{\nusingle / N}} \right),
    \label{eq:pd}
\end{equation}
where $\mathsf{Q(\cdot)}$ is the tail distribution function of the standard normal distribution.
For the $(5,10,L,N)$ SC-LDPC code ensemble, we use $\nusingle \approx 0.424$~\cite{ref:Soko20}.

The approximation~\eqref{eq:pd} assumes perfect coupling of the two sides of a doping point---if one side exhausts degree-one CNs at the local minimum $\ell^*$, so does the other, and no decoding waves emerge.
Conversely, if one side succeeds in producing a decoding wave, the other produces a wave as well.
Moreover, we also use this approximation for the values above the threshold $\estard$, effectively assuming that the probability of doping to work is odd-symmetric around~$(\e\!=\!\estard,1\!-\!\pd\!=\!0.5)$.

\begin{figure}[!t]
    \centering
    \includegraphics[width=\columnwidth]{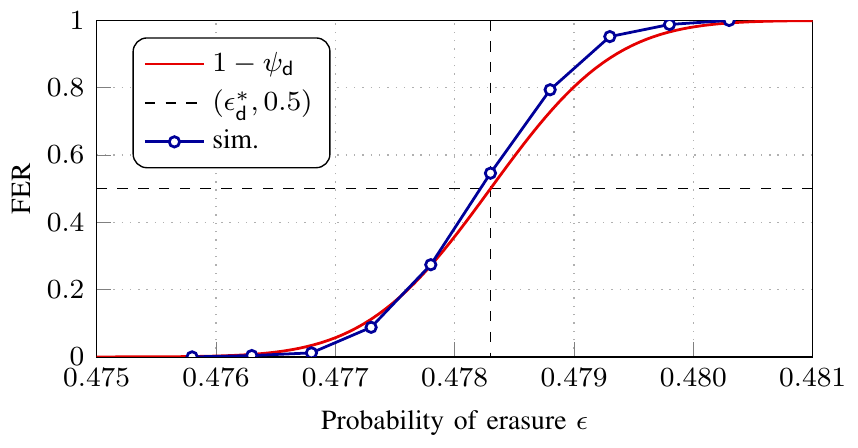}
    \vspace{-15pt}
    \caption{The FER for a tail-biting doped SC-LDPC code ensemble and the corresponding predictions of the probability of doping failure.}
\label{fig:doping_wave}
\vspace{-3ex}
\end{figure}

Fig.~\ref{fig:doping_wave} offers some evidence  of this symmetry.
It shows the simulated FER for the tail-biting doped $(5,10,L\!=\!23,\Pt_3,N\!=\!10^5)$ SC-LDPC code ensemble with $\Di = \{ 0, 1, 2 \}$ (blue curve with circles).
We employ tail-biting to remove the decoding waves from termination boundaries and use a relatively short chain length $L$ to increase the probability that, once formed, the decoding waves from the doping point successfully propagate through the chain, thereby maximizing the influence of doping on the FER.
The red solid curve corresponds to the analytical approximation $1 - \pd$ with $\pd$ given in~\eqref{eq:pd}.
The good agreement between the simulated error rate and the analytical approximation does suggest symmetry around~$(\estard,0.5)$.
\vspace{-5pt}

\subsection{Error Rates for Regularly Doped Streams}
\label{sse:rates}

We are now ready to provide the finite-length scaling law for the error rates in streaming scenarios.
We denote the BER and BLER of the terminated $(\dv,\dc,L,N)$ SC-LDPC code ensemble decoded with SW of size $W$ by $\BERtw^{(L,W)}$ and $\BLERtw^{(L,W)},$ respectively.
Analogously, we denote the corresponding error rates for the semi-infinite ensemble with regular doping every $\Ls$ positions  by $\BERs^{(\Ls,W)}$ and $\BLERs^{(\Ls,W)}.$
We assume that when a doping point works, which happens with probability $\pd$, it equates to full termination.
Otherwise, the system operates as if doping were not applied at all.
As a result of this ``switch-like'' behavior of the doping points, a regularly doped semi-infinite chain effectively splits into a series of terminated chains of different length, a phenomenon that we exploit next.

Consider a doping point that does trigger a decoding wave.
Let $L'$ denote the distance along the chain to the \textit{next} successful doping point, measured in the number of VN-containing spatial positions.
Since doping points trigger decoding waves independently, $L'$ is (up to a scaling by $\Ls$) a geometrically distributed random variable with success probability $\pd$, i.e.,
\begin{equation}
    \operatorname{Pr} \left\{ L' = k\Ls \right\} = (1 - \pd)^{k - 1} \pd.
    \label{eq:pr_lgeom}
\end{equation}
Therefore, we estimate the BER and BLER for the semi-infinite regularly doped ensemble by averaging the error rates for the terminated ensemble over the geometric distribution as
\begin{align}
    P_{\boldsymbol{\cdot},\mathsf{str}}^{(\Ls,W)} &= \sum_{k=1}^{\infty} P_{\boldsymbol{\cdot},\mathsf{t,sw}}^{(k\Ls,W)} (1 - \pd)^{k - 1} \pd \label{eq:er_stream}
\end{align}
with corresponding subscripts substituted by the placeholders.
The component error rates of terminated SC-LDPC code ensembles without doping, i.e., $\BERtw$ and $\BLERtw$, are estimated as in~\cite{ref:Soko20}.
\begin{samepage}
We remark that the approximation in~\eqref{eq:er_stream} ignores possible VN-containing spatial positions within a doping point.

\section{Numerical Results}
\label{sec:results}
\end{samepage}
\begin{figure}
    \centering
    \includegraphics[width=\columnwidth]{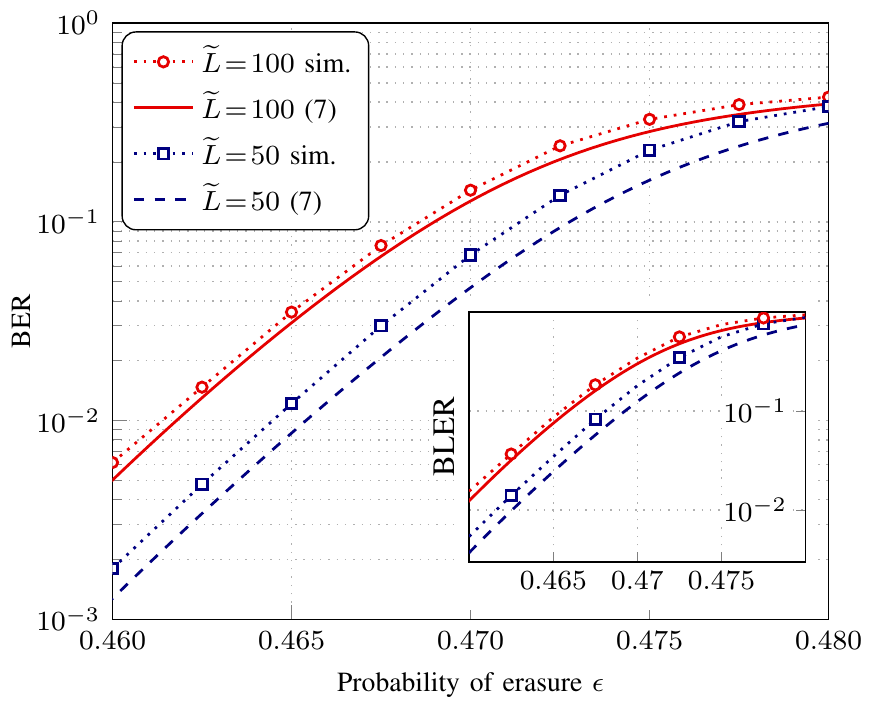}
    \vspace{-15pt}
    \caption{Error rates for semi-infinite $(5,10,\Ls,\Pt_3,N\!=\!10^3)$ SC-LDPC code ensembles with regular doping under SW decoding with $W\!=\!20$.}
\label{fig:er_ber}
\vspace{-3ex}
\end{figure}

Fig.~\ref{fig:er_ber} shows the BER and BLER for the semi-infinite $(5,10,\Ls,\Pt_3,N\!=\!1000)$  ensemble with regular doping $\Di = \{ 0, 1, 2 \}$ spaced by $\Ls=50$ and $\Ls=100$ positions under SW decoding with $W\!=\!20$.
The analytical approximations~\eqref{eq:er_stream} (solid and dashed) are in  good agreement with numerical simulations (dotted curves with markers).
For the simulated error rates, we follow~\cite{ref:Soko20} and ignore all degree-2 stopping sets.
The match improves as we increase the distance between doping points from $\Ls\!=\!50$ (blue) to $\Ls\!=\!100$ (red).

\begin{figure}[t!]
    \centering
    \includegraphics[width=\columnwidth]{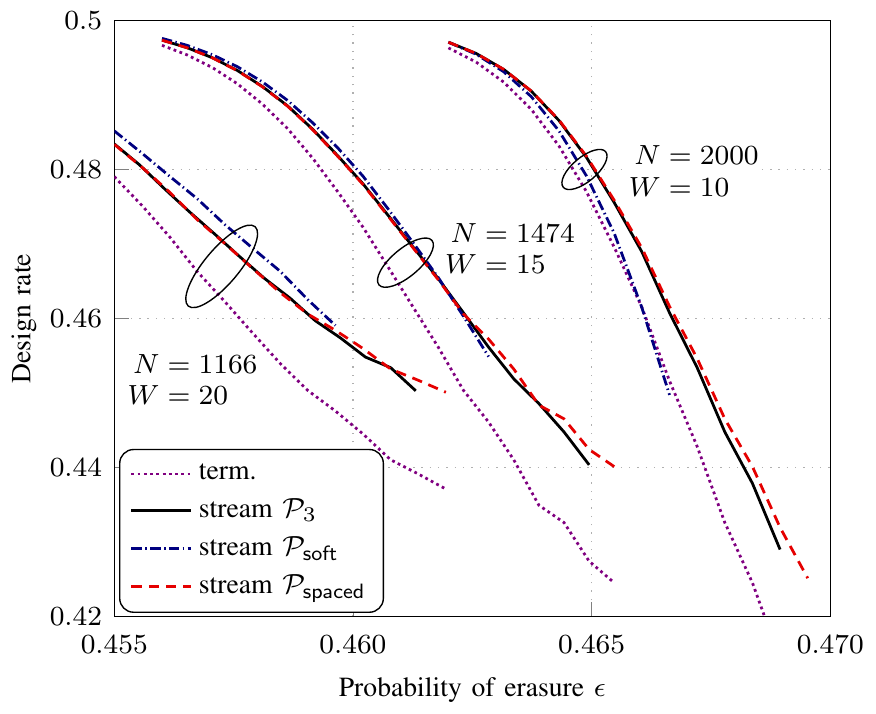}
    \vspace{-15pt}
    \caption{Achievable design rates for semi-infinite $(5,10,\Ls,\Pt,N)$ SC-LDPC code ensembles with regular doping and different doping patterns (Table~\ref{tab:thresholds}).}
\label{fig:achievable_rates}
\vspace{-3ex}
\end{figure}

Fig.~\ref{fig:achievable_rates} compares the achievable design rates associated with different doping patterns under SW decoding for streaming applications.
Specifically, we consider the $(5,10,\Ls,\Pt,N)$ SC-LDPC code ensemble and fix the target BER at $10^{-4}$, focusing on three combinations of $N$ and $W$ that yield the same decoding latency of approximately $28 \cdot 10^3$ bits.
For each $\e$,~we use the finite-length scaling law~\eqref{eq:er_stream} to find the highest $\Ls$ that satisfies $\BERs^{(\Ls,W)} < 10^{-4}$, which in turn determines the highest design rate.
The apparent ``noisiness'' of the resulting curves is due to the fact that we can only increase $\Ls$ in discrete steps; it can be regarded as quantization error.
For each pair $(N,W)$ we compare four doping patterns specified in Table~\ref{tab:thresholds}: full termination (purple dotted), three consecutive doped positions $\Pt_3$ (black solid), soft doping $\Pt_{\mathsf{soft}}$ (blue dash-dotted), and three spaced doped positions $\Pt_{\mathsf{spaced}}$ (red dashed).
We observe that doping can provide higher rates than full termination for the same target BER.
Moreover, soft doping  yields the highest rate for $N\!=\!1166$, $W\!=\!20$ while performing worst for $N\!=\!2000$, $W\!=\!10$, emphasizing the importance of finite-length analysis in parameter optimization.
The proposed scaling law~\eqref{eq:er_stream} can be used to considerably speed up the search for an optimal combination of code parameters---obtaining the results in Fig.~\ref{fig:achievable_rates} via  simulations is computationally challenging.

\vspace{5pt}

\section{Conclusion}
\label{sec:conclusion}

\vspace{-5pt}

We demonstrated that a doping point is well modeled by a probabilistic switch between full termination and a coupled chain with no doping at all.
We used this model to derive a finite-length scaling law that yields accurate approximations of the BER and BLER of semi-infinite SC-LDPC code ensembles with regular doping over the BEC under SW decoding.
The proposed scaling law can be applied to both VN doping proposed by Zhu~\textit{et al.}~\cite{ref:Zhu20_vn} and to optimized shortening by Cammerer~\textit{et al.}~\cite{ref:Camm16}.
We use it to demonstrate that in streaming applications doping can provide higher design rates while guaranteeing the same target BER as periodic full termination.

\end{document}